\documentclass[3p,twocolumn]{elsarticle}

\usepackage{graphicx}

\def\bea{\begin{eqnarray}}
\def\eea{\end{eqnarray}}
\def\be{\begin{equation}}
\def\ee{\end{equation}}

\begin{document}
\begin{frontmatter}
\title{Energy dependence of pion interferometry scales in ultra-relativistic heavy ion collisions}

\author[yuk1,yuk2]{Iu.A.~Karpenko}
\ead{karpenko@bitp.kiev.ua}
\author[yuk1,yus2]{Yu.M.~Sinyukov}
\ead{sinyukov@bitp.kiev.ua}

\address[yuk1]{Bogolyubov Institute for Theoretical
Physics, Metrolohichna str. 14b, 03680 Kiev-143,  Ukraine}
\address[yuk2]{SUBATECH, Universit\'e de Nantes, IN2P3/CNRS EMN, Nantes, France}
\address[yus2]{ExtreMe Matter Institute EMMI, GSI
Helmholtz Zentrum f\"ur
  Schwerionenforschung,
D-64291 Darmstadt, Germany}
\begin{keyword}
A+A collisions, SPS, RHIC, LHC, Interferometry radii, HBT puzzle,
Transverse spectra, Pre-thermal flows, Energy density
\end{keyword}
\begin{abstract}
A study of energy behavior of the pion spectra and interferometry
scales is carried out for the top SPS, RHIC and LHC energies within
the hydrokinetic approach. The latter allows one to describe
evolution of quark-gluon and hadron matter as well as continuous
particle emission from the fluid in agreement with the underlying
kinetic equations. The main mechanisms that lead to the paradoxical,
at first sight, behavior of the interferometry scales, are exposed.
In particular, a slow decrease and apparent saturation of
$R_{out}/R_{side}$ ratio with an energy growth happens due to a
strengthening of positive  correlations between space and time
positions of pions emitted at the radial periphery of the system.
Such en effect is a consequence of the two factors: a developing of
the pre-thermal collective transverse flows and an increase of the
initial energy density in the fireball.
\end{abstract}
\end{frontmatter}

\section{Introduction}

The pion interferometry method, which is based on the Bose-Einstein
or Fermi-Dirac interference of identical particles, has been
proposed first in Refs. \cite{Kopylov,Cocconi} for measurements of
the geometrical sizes and shapes of the interaction region in
hadronic collisions. Then it has been developed in Refs.
\cite{Pratt,MakSin, Averch, Hama} as a tool for a study of rapidly
expanding fireballs formed in ultra-relativistic heavy ion
collisions. Despite the extremely small sizes of such systems - the
order of the value  is around 10$^{-14}$ m - they have,
nevertheless, a pronounced inhomogeneous structure. The generalized
treatment of the interferometry measurements asserts that the
measured scales - the interferometry radii - are associated just
with the homogeneity lengths in the system \cite{Sin}. Only in the
very particular case of a finite homogeneous system such lengths
correspond to the total geometrical sizes, but normally they are
smaller than the latter. The interferometry scanning of
femto-systems at various total momenta of pion pairs allows one to
analyze the homogeneity lengths related to different space-time
regions of the expanding fireball \cite{Averch, AkkSin}. An
understanding of interferometry in terms of the homogeneity length,
as opposed to the simple-mind geometrical picture, provides
explanation to some, at first sight paradoxical, results at RHIC.

Naively it was expected that when the energy of colliding nuclei
increases, the pion interferometry volume $V_{int}$ - product of the
interferometry radii in three orthogonal directions - rises at the
same maximal centrality for Pb+Pb and Au+Au collisions just
proportionally to $\frac{dN_{\pi}}{dy}$. However, when experiments
at RHIC starts, an increase of the interferometry volume with energy
turns out to be essentially flatter then the proportionality low.
This is one of the components of so-called RHIC HBT puzzle
\cite{HBTpuzzle}.

Another item of the HBT puzzle is an unexpected
RHIC result as for the ratio of the two transverse interferometry
radii: one that is measured in the direction of the sum of
transverse momenta of the (pion) pair, $R_{out}$, and another,
$R_{side}$, - orthogonally to it and the beam axis. While the latter
is associated with transverse homogeneity length, the former
includes besides that also additional contributions, in particular,
the one which is related to a duration of the pion emission. Since
the lifetime of the systems obviously should grow with collision
energy, if it is accompanied by an increase of the initial energy
density and/or by a softening of the equation of state due to phase
transition between hadron matter and quark-gluon plasma (QGP), then
the duration of pion emission should grow with energy and
$R_{out}/R_{side}$ ratio could increase. A very large value of the
ratio was predicted as a signal of the QGP formation \cite{Bertsch}.
The RHIC experiments brought the result: the ratio
$R_{out}/R_{side}\approx 1$ and similar or even smaller than at SPS.

In this letter we analyze the pion spectra and the interferometry
scales as depending on the energy of central Pb+Pb and Au+Au
collisions supposing the formation of QGP. The aims of the study are
to describe these pion observables at the top SPS and RHIC energies
and make predictions for LHC energies, also to understand the
physical mechanisms responsible for the peculiarities of energy
dependence of the interferometry radii, in particular, the
$R_{out}/R_{side}$ ratio. Numerical calculations are based on the
HydroKinetic Model (HKM) \cite{PRL,PRC} that allows one to describe
the evolution of quark-gluon and hadron matter and continuous
particle emission from the fluid in agreement with the underlying
kinetic equations.

\section{Hydro-kinetic approach to A+A collisions}

Let us briefly describe the main features of the HKM \cite{PRL,PRC}.
It incorporates hydrodynamical expansion of the systems formed in A
+ A collisions and their dynamical decoupling described by escape
probabilities. The method corresponds to a generalized relaxation
time approximation for the particle emission functions applied to
inhomogeneous systems expanding into vacuum according to the
Boltzmann equations.  This method also allows one to describe the
viscous effects at the hadronic stage of the evolution. The basic
hydro-kinetic code, proposed in \cite{PRC}, is modified now to
include decays of resonances into expanding hadronic non-chemically
equilibrated system and, based on the resulting composition of the
hadron-resonance gas at each space-time point, to calculate the
equation of state (EoS) in a vicinity of this point. The obtained
local EoS allows one to determine the further evolution of the
considered fluid elements. The complete picture of the physical
processes in central Pb + Pb and Au + Au collisions encoded in
calculations is the following.

\subsection{Initial conditions}

 Our results are all related to the
central rapidity slice where we use the boost-invariant Bjorken-like
initial condition. We consider the proper time of thermalization of
quark-gluon matter as  the minimal one discussed in the literature,
$\tau_0=1$ fm/c. The initial energy density in the transverse plane
is supposed to be Glauber-like \cite{Kolb}, i.e. is proportional to
the participant nucleon density  for Pb+Pb (SPS) and Au+Au (RHIC,
LHC) collisions with zero impact parameter. The height of the
distribution - the maximal initial energy density -
$\epsilon(r=0)=\epsilon_0$ is the fitting parameter. From analysis
of pion transverse spectra we choose it for the top SPS energy to be
$\epsilon_0 = 9$ GeV/fm$^3$ ($\langle\epsilon\rangle_0$ = 6.4
GeV/fm$^3$), for the top RHIC energy  $\epsilon_0 = 16.5$ GeV/fm$^3$
($\langle\epsilon\rangle_0$ = 11.6 GeV/fm$^3$). The brackets $<...>$
correspond to mean value over the distribution associated with the
Glauber transverse profile. We also demonstrate results at
$\epsilon_0 = 40$ GeV/fm$^3$ and $\epsilon_0 = 60$ GeV/fm$^3$ that
probably can correspond to the LHC energies of 4 and 5.6 ATeV. As
for the latter, according to \cite{Lappi}, at the top LHC energy
$\epsilon_{max}$ recalculated to the time 1 fm/c is $0.07*700=49$
GeV/fm$^3$. We suppose that soon after thermalization the matter
created in A+A collision at RHIC energies is in the quark-gluon
plasma (QGP) state. Also, at time $\tau_i$, there is peripheral
region with relatively small initial energy densities: $\epsilon(r)<
0.5$ GeV/fm$^3$. This part of the matter ("corona") does not
transform into QGP and have no chance to be involved in
thermalization process \cite{Werner}. By itself the corona gives no
essential contribution to the hadron spectra \cite{Werner}.This part
of the matter, probably, is not thermalized. One should consider it
separately from the thermal bulk of the matter and should not
include in hydrodynamic evolution.

At the time of thermalization, $\tau_0=1$ fm/c, the system already
has developed collective transverse velocities \cite{flow,JPG}. The
initial transverse rapidity profile is supposed to be linear in
radius $r_T$:
\begin{equation}
y_T=\alpha\frac{r_T}{R_T} \quad where \quad R_T=\sqrt{<r_T^2>}.
\label{yT},
\end{equation}
Here $\alpha$ is the second fitting parameter. Note that the fitting
parameter $\alpha$ should include also a positive correction for
underestimated resulting transverse flow since in this work we do
not take directly into account the viscosity effects \cite{Teaney}
neither at QGP stage nor at hadronic one. In formalism of HKM
\cite{PRC} the viscosity effects at hadronic stage are incorporated
in the mechanisms of the back reaction of particle emission on
hydrodynamic evolution which we ignore in current calculations.
Since the corrections to transverse flows which depend on unknown
viscosity coefficients are unknown, we use fitting parameter
$\alpha$ to describe the "additional unknown portion" of flows,
caused both factors: by a developing of the pre-thermal flows and
the viscosity effects in quark-gluon plasma. The best fits of the
pion transverse spectra at SPS and RHIC are provided at
$\alpha=0.194$ ($\langle v_T \rangle= 0.178$) for SPS energies and
$\alpha=0.28$ ($\langle v_T\rangle=0.25$) for RHIC ones. The latter
value we use also for LHC energies aiming to analyze just influence
of energy density increase. From a fitting of the pion spectra we
can conclude that an "additional portion" of flow is bigger at RHIC
than at SPS where the parameter $\alpha$ is compatible with that
free streaming of partons brings \cite{flow,JPG}. One can conclude,
thus, that longer viscous QGP evolution at RHIC has a bigger impact
for a developing of transverse flows than at the SPS energies.

\subsection{Equation of state}

Following Ref. \cite{PRC} we use at high temperatures the
EoS \cite{Laine} adjusted to the QCD lattice data with the baryonic
chemical potential $\mu_B =0$ and matched with chemically
equilibrated multi-component hadron resonance gas at $T=175$ MeV.
Such an EoS could be a good approximation for the RHIC and LHC
energies; as for the SPS energies we utilize it just  to demonstrate
the energy-dependent mechanism of formation of the space-time
scales. We suppose the chemical freeze-out for the hadron gas at
$T_{ch}=165$ MeV \cite{PBM1}. It guarantees us the correct particle
number ratios for all quasi-stable particles (here we calculate only
pion observables) at least for RHIC. Below $T_{ch}$ a composition of
the hadron gas is changed only due to resonance decays into
expanding fluid. We include 359 hadron states made of u, d, s quarks
with masses up to 2.6 GeV. The EoS in this non-chemically
equilibrated system depends now on particle number densities $n_i$
of all the 359 particle species $i$: $p=p(\epsilon,\{n_i\})$. Since
the energy densities in expanding system do not directly correlate
with resonance decays, all the variables in the EoS depend on
space-time points and so an evaluation of the EoS is incorporated in
the hydrodynamic code. We calculate the EoS below $T_{ch}$ in the
Boltzmann approximation of ideal multi-component hadron gas.

\subsection{Evolution}

At the temperatures higher than $T_{ch}$ the hydrodynamic evolution
is related to the quark-gluon and hadron phases which are in
chemical equilibrium with zero baryonic chemical potential. The
evolution is described by the conservation law for the
energy-momentum tensor of perfect fluid:
\begin{equation}
\partial_\nu T^{\mu\nu}(x)=0
\label{conservation}
\end{equation}
At $T<T_{ch}$=165 MeV the system evolves as non-chemically
equilibrated hadronic gas. The concept of the chemical freeze-out
implies that afterwards  only elastic collisions and resonance
decays take place because of relatively small densities allied with
a fast rate of expansion at the last stage. Thus, in addition to
(\ref{conservation}), the equations accounting the particle number
conservation and resonance decays are added. If one neglects the
thermal motion of heavy resonances the equations for particle
densities $n_i(x)$ take the form:
\begin{equation}
    \partial_\mu(n_i(x) u^\mu(x))=-\Gamma_i n_i(x) + \sum\limits_j b_{ij}\Gamma_j
    n_j(x)
    \label{paricle_number_conservation}
\end{equation}
where $b_{ij}=B_{ij}N_{ij}$ denote the average number of $i$th
particles coming from arbitrary decay of $j$th resonance,
$B_{ij}=\Gamma_{ij}/\Gamma_{j,tot}$ is branching ratio, $N_{ij}$ is
a number of $i$th particles produced in $j\rightarrow i$ decay
channel. We also can account for recombination in the processes of
resonance decays into expanding medium just by utilizing the
effective decay width $\Gamma_{i,eff}=\gamma\Gamma_i$.  We use
$\gamma = 0.75$ supposing thus that near 30\% of resonances are
recombining during the evolution. All equations (\ref{conservation})
and 359 equations (\ref{paricle_number_conservation}) are solved
simultaneously with calculation of the  EoS,
$p(x)=p(\epsilon(x),\{n_i(x)\})$, at each point $x$.

\subsection{System's decoupling and spectra formation}

During the matter evolution, in fact, at $T\leq T_{ch}$, hadrons
continuously leave the system. Such a process is described by means
of the emission function $S(x,p)$ which is expressed for pions
through the {\it gain} term, $G_{\pi}(x,p)$, in Boltzmann equations
and
the escape probabilities ${\cal P}_{\pi}(x,p)=\exp(-\int\limits_{t}^{\infty}%
dsR_{\pi+h}(s,{\bf r}+\frac{{\bf p}}{p^0}(s-t),p))$:
$S_{\pi}(x,p)=G_{\pi}(x,p){\cal P}_{\pi}(x,p)$ \cite{PRL,PRC}. For
pion emission in relaxation time approximation $G_{\pi}\approx
f_{\pi}R_{\pi+h}+G_{H\rightarrow\pi}$ where $f_{\pi}(x,p)$ is the
pion Bose-Einstein phase-space distribution, $R_{\pi+h}(x,p)$ is the
total collision rate of the pion, carrying momentum $p$, with all
the hadrons $h$ in the system in a vicinity of point $x$,  the term
$G_{H\rightarrow\pi}$ describes an inflow of the pions into
phase-space point $(x,p)$ due to the resonance decays. It is
calculated according to the kinematics of decays with simplification
that the spectral function of the resonance $H$ is
$\delta(p^2-\langle m_H\rangle^2)$. The cross-sections in the
hadronic gas, that determine via the collision rate $R_{\pi+h}$ the
escape probabilities ${\cal P}(x,p)$ and emission function $S(x,p)$,
are calculated in accordance with the UrQMD method \cite{UrQMD}. The
spectra and correlation functions are found from the emission
function $S$ in the standard way (see, e.g., \cite{PRL}).

\section{Results and conclusions}
The pion emission function per unit (central) rapidity, integrated
over azimuthal angular and transverse momenta, is presented in Fig.
1 for the top SPS, RHIC and LHC energies as a function of transverse
radius $r$ and proper time $\tau$.  The two fitting parameters
$\epsilon_0$ and $\langle v_T \rangle$ are fixed as discussed above
and marked in figures. The pion transverse momentum spectrum, its
slope as well as the absolute value, and the interferometry radii,
including $R_{out}$ to $R_{side}$ ratio, are in a good agreement
with the experimental data both for the top SPS and RHIC energies.

As one can see particle emission lasts a total lifetime of the
fireballs; in the central part, ${\bf r}\approx 0$, the duration is
half of the lifetime.  Nevertheless, according to very recent
results \cite{PRC, freeze-out}, the Landau/Cooper-Frye presentation
of sudden freeze-out could be applied in a generalized form
accounting for momentum dependence of the freeze-out hypersurface
$\sigma_p(x)$; now $\sigma_p(x)$ corresponds to the {\it maximum of
emission function} $S(t_{\sigma}({\bf r},p),{\bf r},p)$ at fixed
momentum ${\bf p}$ in an appropriate region of ${\bf r}$. This
finding gives one possibility to keep in mind the known results
based on the Cooper-Frye formalism, applying them to a surface of
the maximal emission for given $p$. Then the typical features of the
energy dependence can be understood as follows. The inverse of the
spectra slopes, $T_{eff}$, grows with energy, since as one sees from
the emission functions, the duration of expansion increases with
initial energy density and, therefore, the fluid element gets more
transverse collective velocities $v_T$ when reach a decoupling
energy densities. Therefore the blue shift of the spectra becomes
stronger. A rise of the transverse collective flow with energy leads
to some compensation of an increase of $R_{side}$: qualitatively the
homogeneity length at decoupling stage is $R_{side}=
R_{Geom}/\sqrt{1+\langle v_{T}^2\rangle m_{T}/2T}$, (see, e.g.,
\cite{AkkSin}). So, despite the significant increase of the
transverse emission region, $R_{Geom}$, seen in Fig.1, a
magnification of collective flow partially compensates this. It
leads to only a moderate increase of the $R_{side}$ with energy.
Since the temperatures in the regions of the maximal emission
decrease very slowly when initial energy density grows (e.g., the
temperatures for SPS, RHIC and LHC are correspondingly 0.105, 0.103
and 0.95 MeV  for $p_T=0.3$ GeV/c ) the $R_{long}\sim
\tau\sqrt{T/m_T}$ \cite{Averch} grows proportionally to an increase
of the proper time associated with the hypersurface
$\sigma_{p_T}(x)$ of {\it maximal} emission. As we see from Fig. 1
this time grows quite moderate with the collision energy.

\begin{figure*}[htb]
\centering
 \includegraphics[scale=0.3]{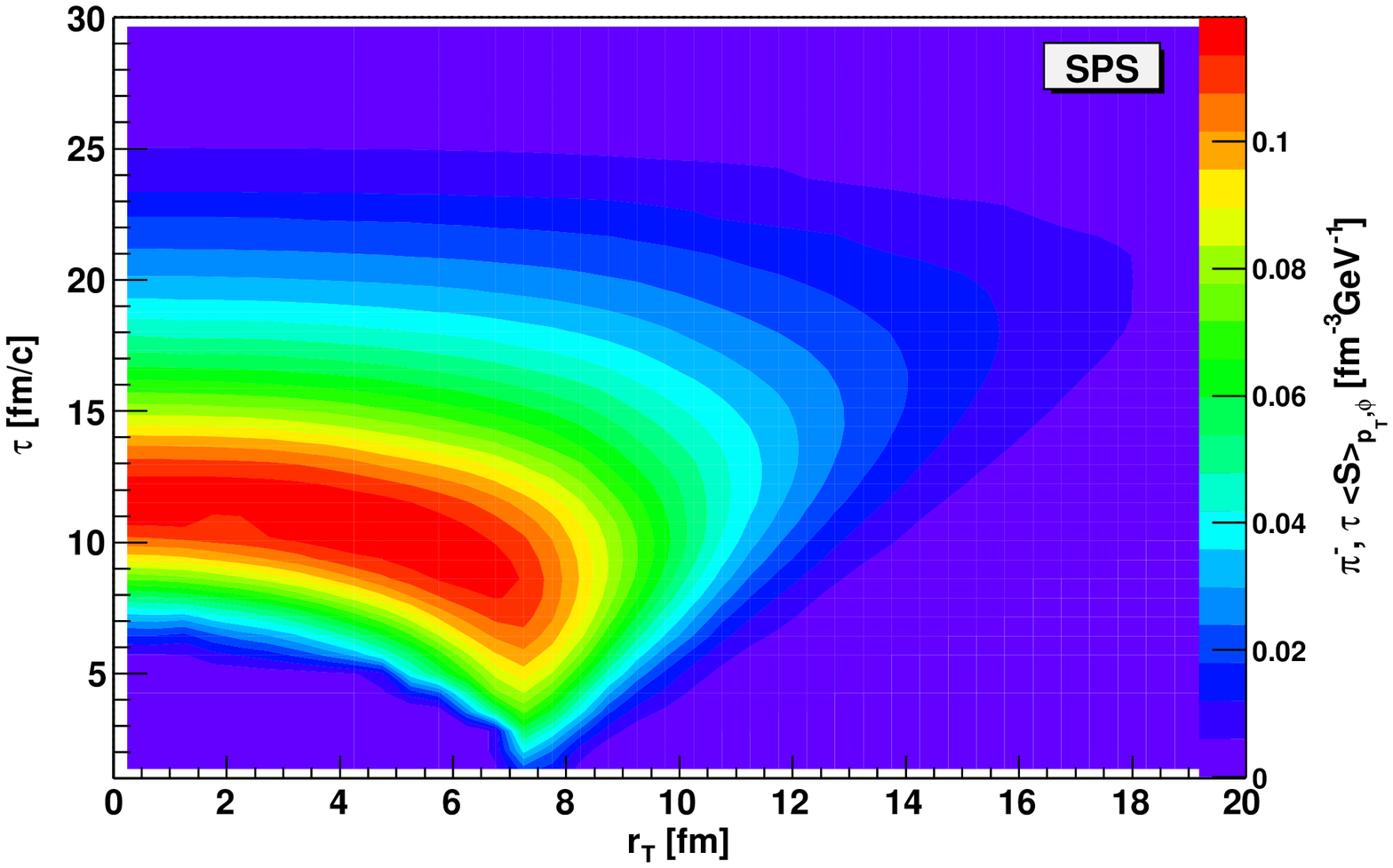}%
 \includegraphics[scale=0.3]{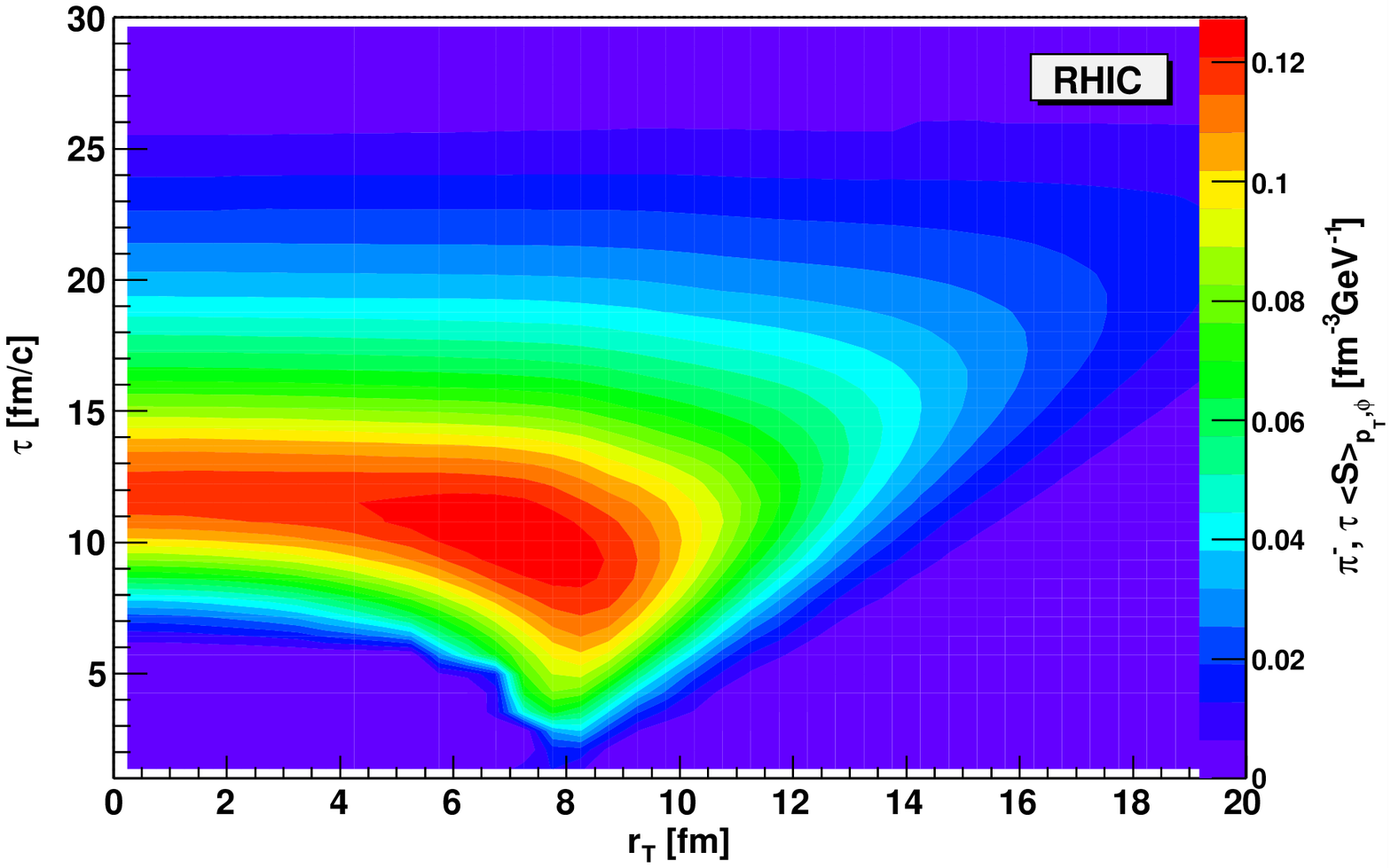}%
 \includegraphics[scale=0.3]{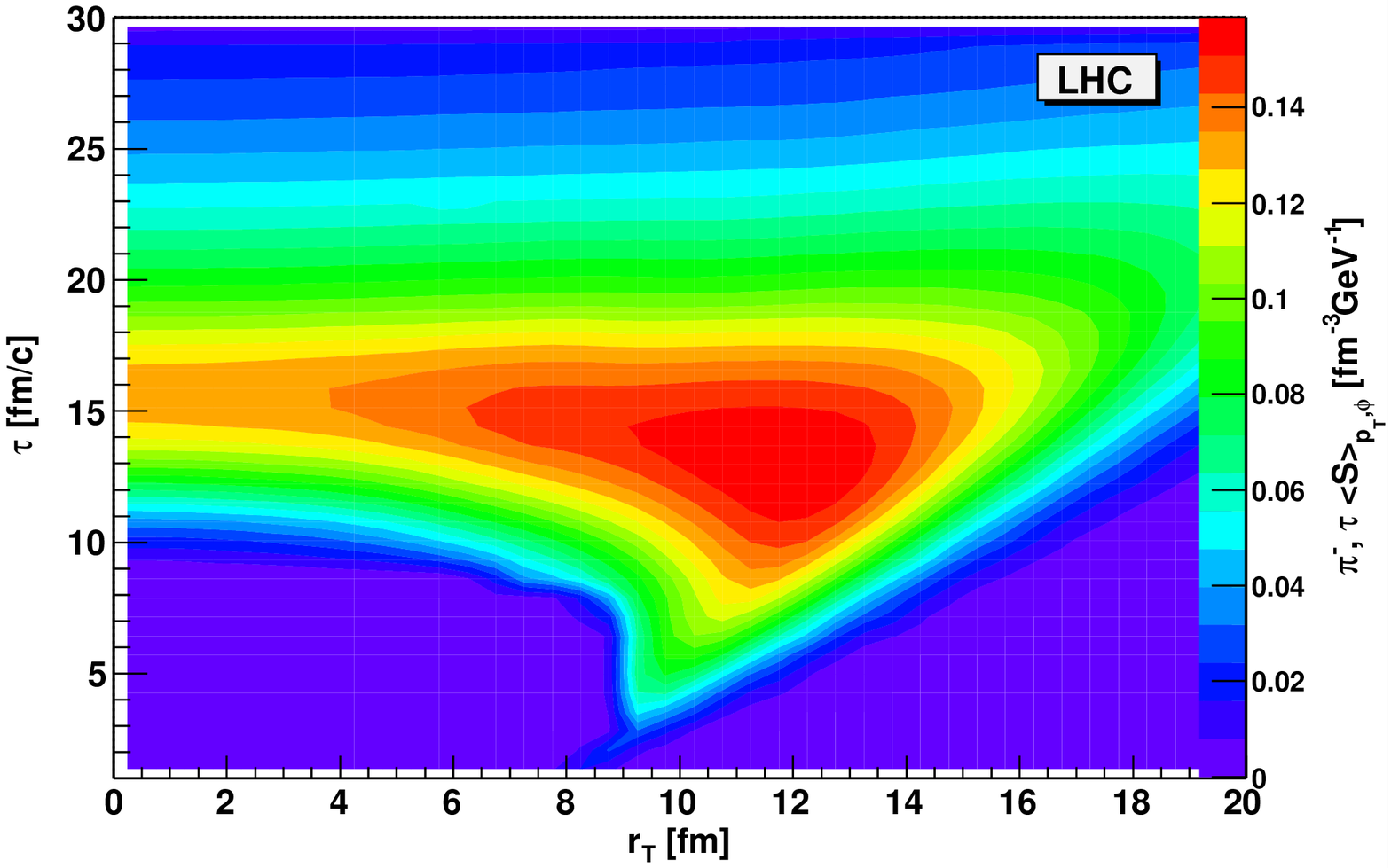}
 \includegraphics[scale=0.3]{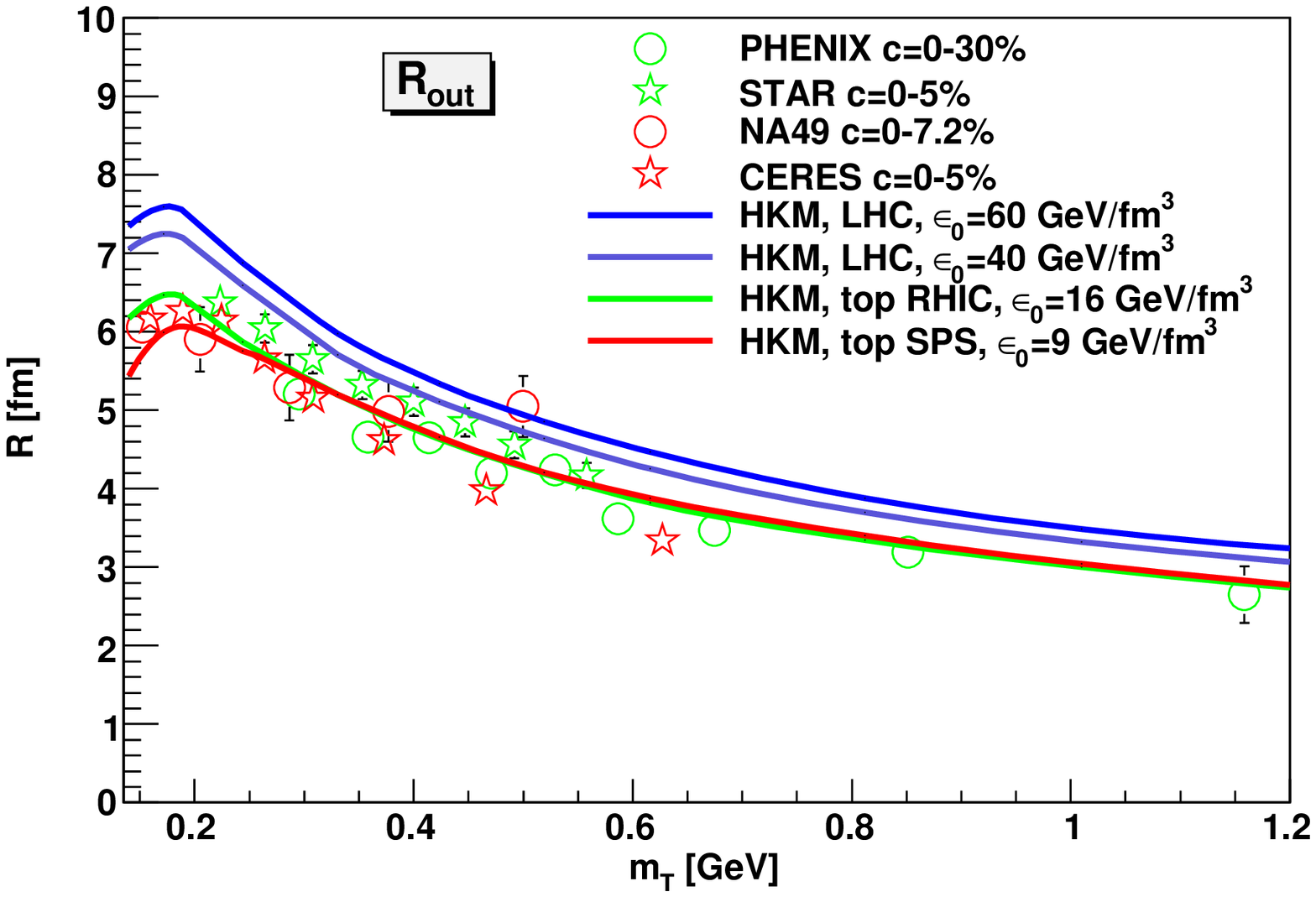}%
 \includegraphics[scale=0.3]{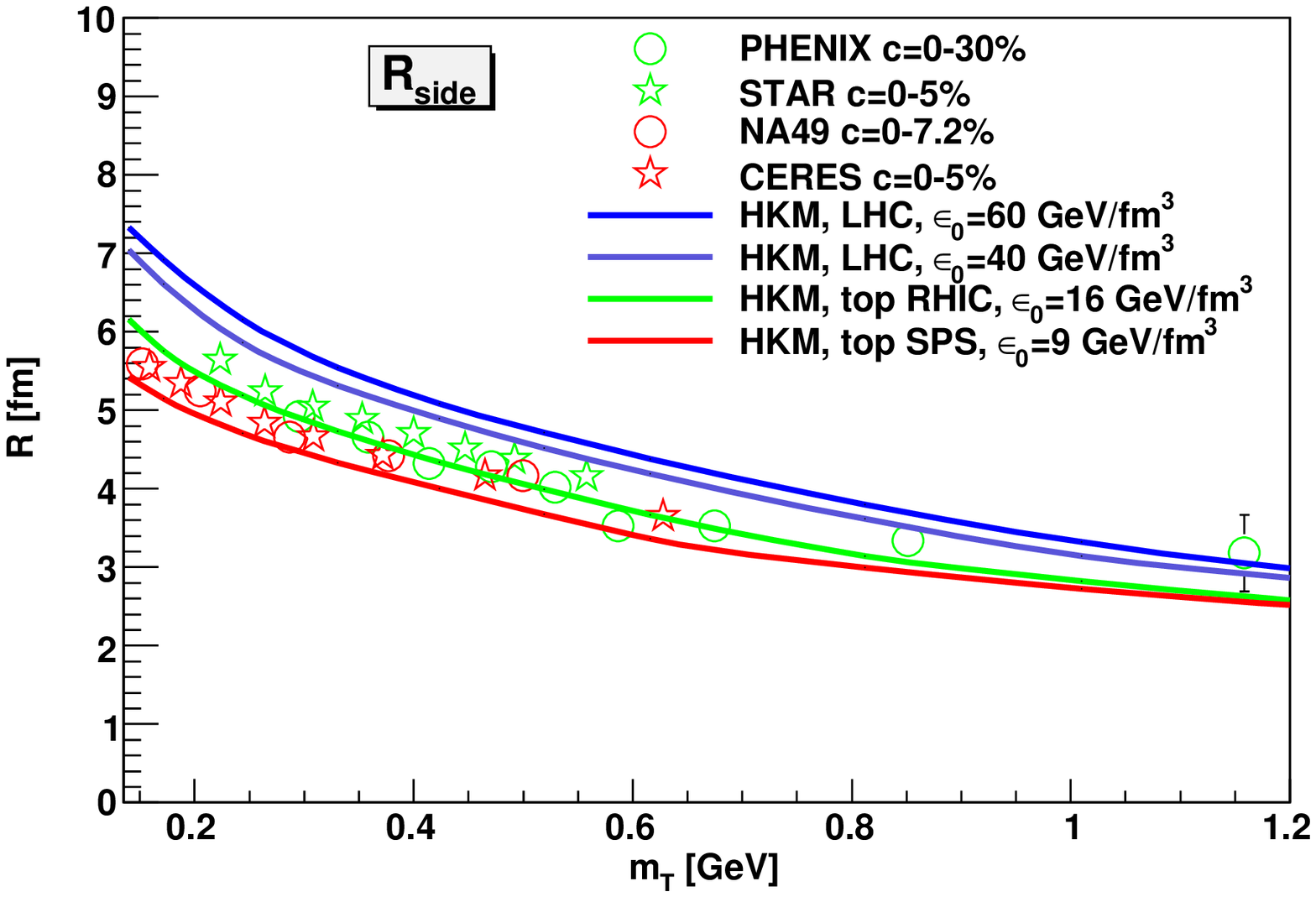}%
 \includegraphics[scale=0.3]{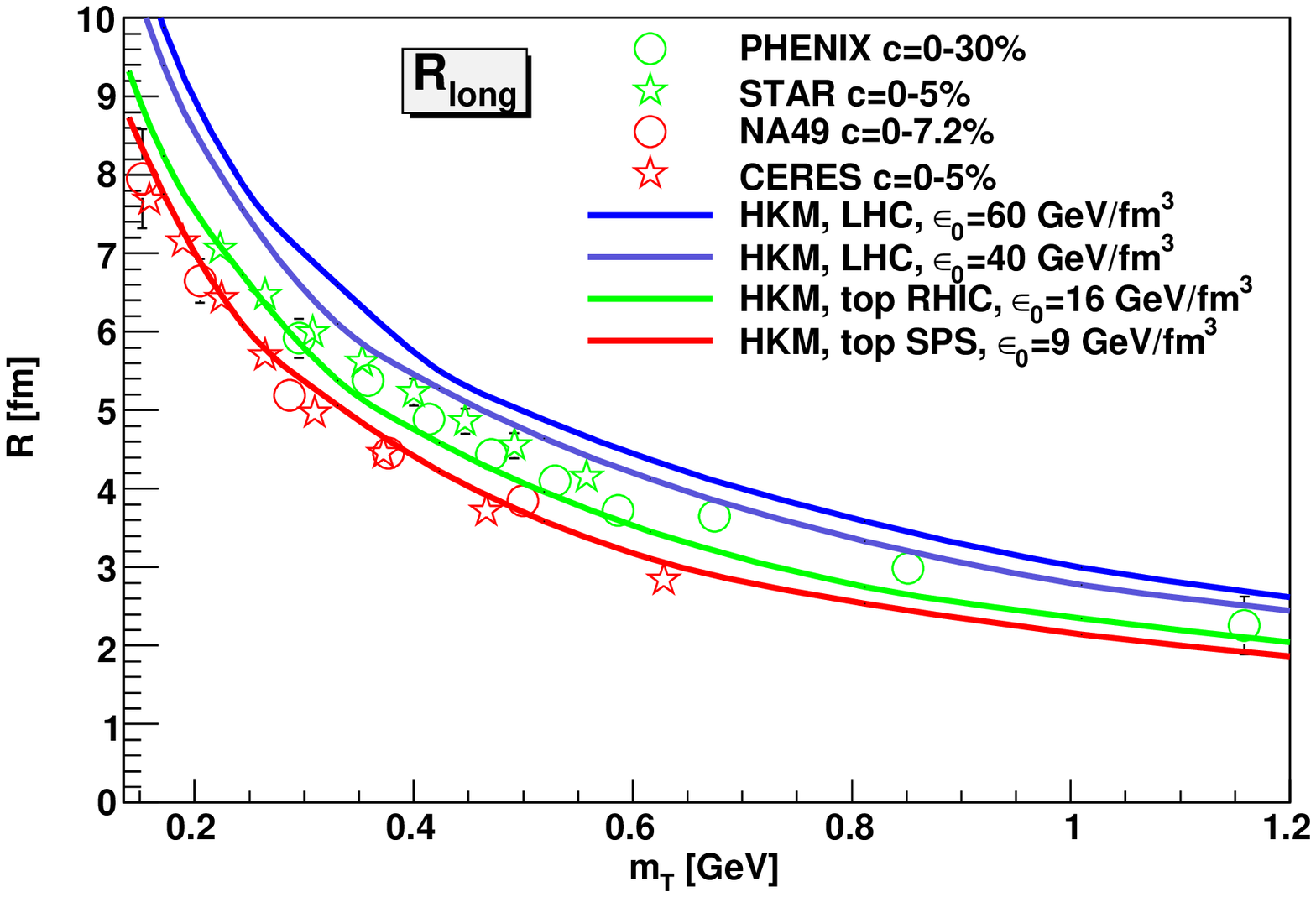}
 \includegraphics[scale=0.3]{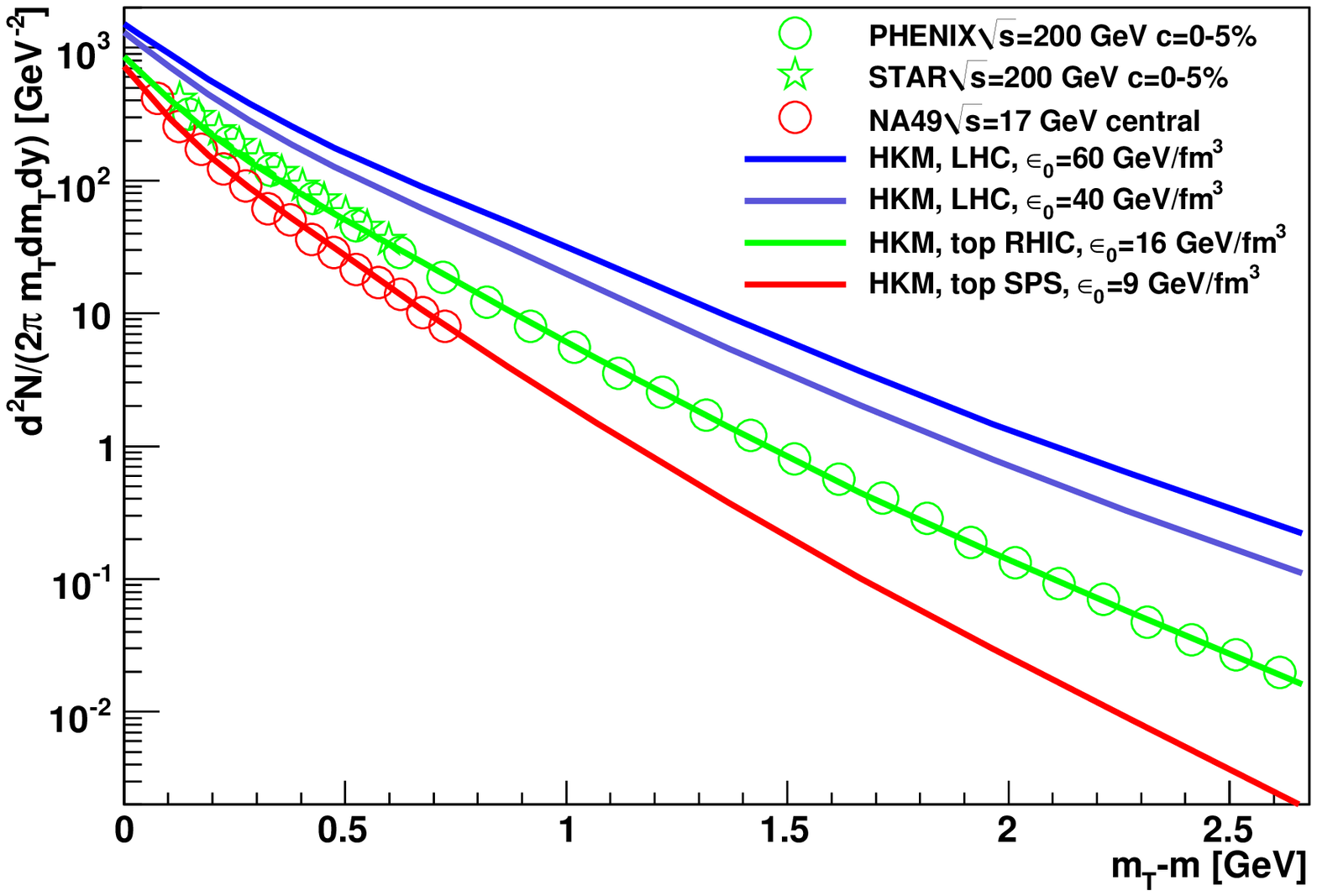}%
 \includegraphics[scale=0.3]{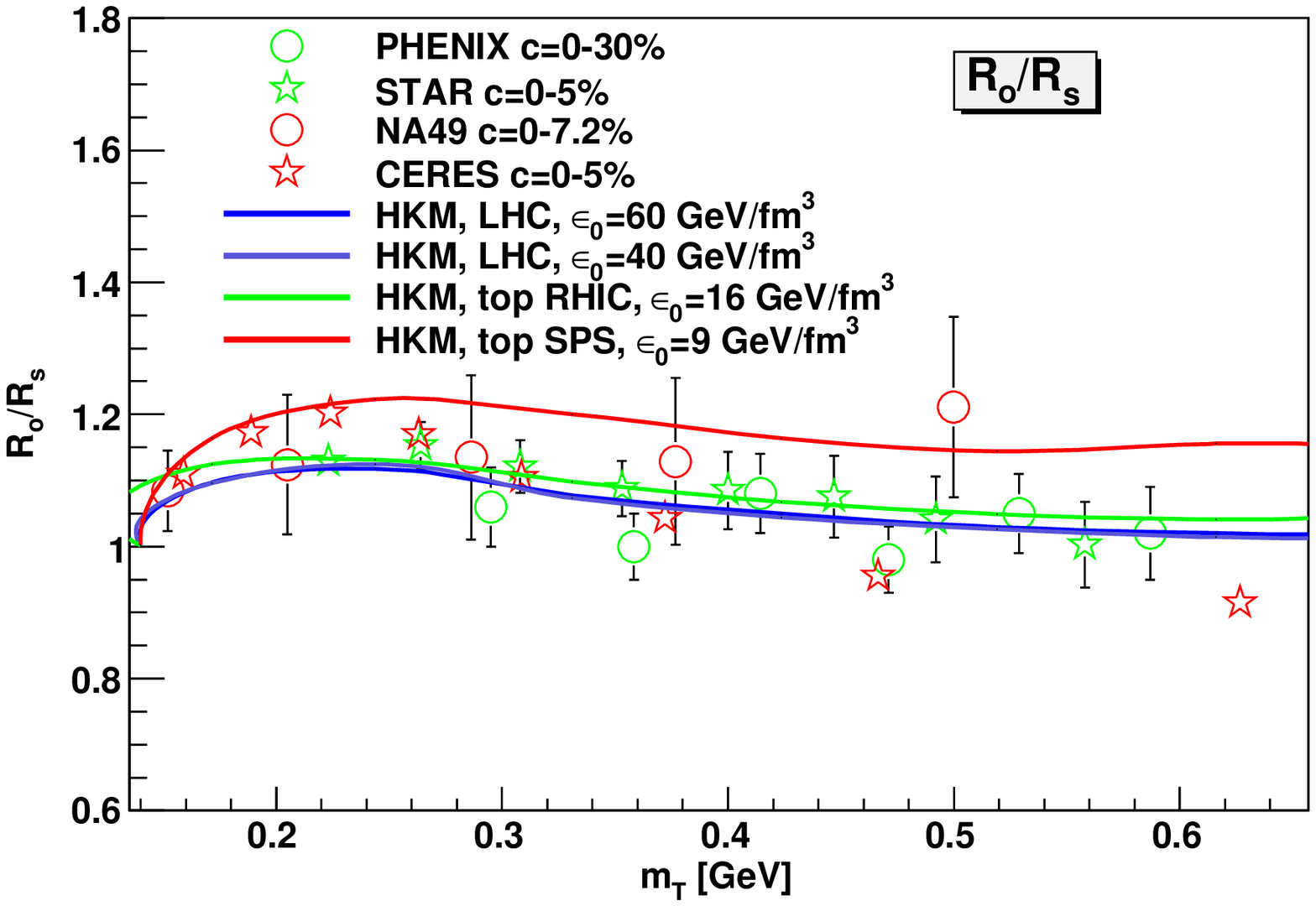}

 \vspace{-0.1in}
\caption{The $p_T$-integrated emission functions of negative pions
for the top SPS, RHIC and LHC energies (top); the interferometry
radii (middle) $R_{out}/R_{side}$ ratio and transverse momentum
spectra (bottom) of negative pions  at different energy densities,
all calculated in HKM model. The experimental data are taken from
CERES \cite{ceres} and NA-49 Collaborations \cite{na49-spectra,
na49-hbt} (SPS CERN), STAR \cite{star-spectra, star-hbt} and PHENIX
\cite{phenix-spectra, phenix-hbt} Collaborations (RHIC BNL)}
\end{figure*}

A non-trivial result concerns the energy behavior of the
$R_{out}/R_{side}$ ratio. It slowly drops when energy grows and
apparently is saturated at fairly high energies at the value close
to unity (Fig.1). To clarify the physical reason of it let us make a
simple half-quantitative analysis. As one can see in Fig. 1, the
hypersurface of the maximal emission can be approximated as
consisting of two parts: the "volume" emission ($V$) at $\tau
\approx const$ and "surface" emission ($S$). A similar picture
within the Cooper-Frye prescription, which generalizes the
blast-wave model \cite{blast-wave} by means of including of the
surface emission has been considered in Ref. \cite{Marina}. If the
hypersurface of maximal emission $\tilde{\tau}(r)$ is double-valued
function, as in our case, then at some transverse momentum $p_T$ the
transverse spectra and HBT radii will be formed mostly by the two
contributions from the different regions with the homogeneity
lengths $\lambda_{i,V}=\sqrt{<(\Delta r_i)^2>}$ ($i$ = side, out) at
the $V$-hypersurface and with the homogeneity lengths
$\lambda_{i,S}$ at the S-hypersurface. Similar to Ref.\cite{AkkSin},
one can apply at $m_T/T\gg1$ the saddle point method when calculate
the single and two particle spectra using the boost-invariant
measures $\mu_V=d\sigma^V_{\mu}p^{\mu}= \widetilde{\tau}(r)r dr
d\phi d\eta
(m_T\cosh(\eta-y)-p_T\frac{d\widetilde{\tau}(r)}{dr}\cos(\phi -
\alpha))$ and $\mu_S=d\sigma^S_{\mu}p^{\mu}= \widetilde{r}(\tau)
\tau d\tau d\phi d\eta
(-m_T\cosh(\eta-y)\frac{d\widetilde{r}(\tau)}{d\tau}+p_T\cos(\phi -
\alpha))$ for $V$- and $S$- parts of freeze-out hypersurface
correspondingly (here $\eta$ and $y$ are space-time and particle
pair rapidities, the similar correspondence is for angles $\phi$ and
$\alpha$, also note that
$\frac{p_T}{m_T}>\frac{d\widetilde{r}(\tau)}{d\tau}$ \cite{PRC,
freeze-out}). Then one can write, ignoring for simplicity the
interference (cross-terms) between the surface and volume
contributions,
\begin{eqnarray}
R_{side}^2 = c_V^2\lambda_{side,V}^2+c_S^2\lambda_{side,S}^2 \label{3} \\
R_{out}^2 = c_V^2\lambda_{out,V}^2+c_S^2\lambda_{out,S}^2(1-
\frac{d\tilde{r}}{d\tau})^2, \label{4}
\end{eqnarray}
where the coefficients $c_V^2+c_S^2\leq1$ and we take into account
that at $p^0/T\gg1$ for pions $\beta_{out}=p_{out}/p^0 \approx 1$.
All homogeneity lengths depend on mean transverse momentum of the
pion pairs $p_T$. The slope $\frac{d\tilde{r}}{d\tau}$ in the region
of homogeneity expresses the strength of $r-\tau$ correlations
between the space and time points of particle emission at the
$S$-hypersurface $\tilde{r}(\tau)$. The picture of emission in Fig.
1 shows that when the energy grows the correlations between the time
and radial points of the emission become positive,
$\frac{d\tilde{r}}{d\tau}> 0$, and they increase with energy
density. The positivity is caused by the initial radial flows
\cite{flow} $u^r(\tau_0)$, which are developed at the pre-thermal
stage, and the strengthening of the $r-\tau$ correlations happens
because the non-central $i$th fluid elements, which produce after
their expansion the surface emission, need more time
$\tau_i(\epsilon_0)$ to reach the decoupling density if they
initially have higher energy density $\epsilon_0$. (Let us
characterize this effect by the parameter
$\kappa=\frac{d\tau_i(\epsilon_0)}{d\epsilon_{0}} > 0 $). Then the
fluid elements before their decays run up to larger radial
freeze-out position $r_i$: if $a$ is the average Lorentz-invariant
acceleration of those fluid elements during the system expansion,
then roughly for $i$th fluid elements which decays at time $ \tau_i$
we have at $a\tau_i \gg 1$: $r_i(\tau_i)\approx
r_i(\tau_0)+\tau_i+(u_i^r(\tau_0)-1)/a$. Then the level of $r-\tau$
correlations within the homogeneous freeze-out "surface" region,
which is formed by the expanding matter that initially at $\tau_0$
occupies the region between the transversal radii $r_1(\tau_0)$ and
$r_2(\tau_0)>r_1(\tau_0)$, is
\begin{equation}
 \frac{d\tilde{r}}{d\tau} \approx \frac{r_1(\tau_1)-r_2(\tau_2)}{\tau_1-\tau_2}
 \approx 1-\frac{R}{\epsilon_0\kappa}\label{5}
\end{equation}
and, therefore, the strength of $r-\tau$ correlations grows with
energy: $\frac{d\tilde{\tau}}{dr}\rightarrow 1$. Note that here we
account for $\tau_2 - \tau_1 \approx \kappa(\epsilon_0(r_2(\tau_0))
- \epsilon_0(r_1(\tau_0)))$ and that
$\frac{d\epsilon_0(r)}{dr}\approx -\frac{\epsilon_0}{R}$ where
$\epsilon_0\equiv\epsilon_0(r=0)$ and $R$ is radius of nuclear. As a
result the second S-term in Eq. (\ref{4}) tends to zero at large
$\epsilon_0$ , reducing, therefore, the $R_{out}/R_{side}$ ratio. In
particular, if $\lambda_{side,V}^2 \gg \lambda_{side,S}^2$ then,
accounting for a similarity of the volume emission in our
approximation and in the blast wave model, where as known
$\lambda_{side,V} \approx \lambda_{out,V}$, one can get:
$\frac{R_{out}}{R_{side}}\approx 1 + const\cdot
\frac{R}{\epsilon_0\kappa}\rightarrow 1$ at $\epsilon_0 \rightarrow
\infty$. It is worth noting that  measure $\mu_S$ also tends to zero
when $\frac{d\tilde{\tau}}{dr}\rightarrow 1$ that again reduces the
surface contribution to $side-$ and $out-$ radii at large $p_T$.

The presented qualitative, in fact, analysis demonstrates the main
mechanisms leading to the non-trivial behavior of $R_{out}$ to
$R_{side}$ ratio exposed in detailed HKM calculations, see Fig.1
(bottom, right).

\section{Summary} We conclude that the energy behavior of the pion
interferometry scales and transverse spectra can be understood if
they are analyzed within fairly developed hydro-kinetic models. The
latter should be based on EoS which accounts for a crossover
transition between quark-gluon and hadron matters at high collision
energies and non-chemically equilibrated expansion of the
hadron-resonance gas at the later stage. The process of particle
emission from expanding fireball, that is not sudden and lasts about
system's lifetime, should be correctly treated. The pre-thermal
formation of transverse flow have to be taken into account. Then the
main mechanisms that lead to the paradoxical behavior of the
interferometry scales find a natural explanation. In particular, a
slow decrease and apparent saturation of $R_{out}/R_{side}$ ratio
around unity at high energy happens due to a strengthening  of
positive correlations between space and time positions of pions
emitted at the radial periphery of the system. Such an effect is a
consequence of the two factors accompanying an increase of collision
energy: a developing of the pre-thermal collective transverse flows
and an increase of initial energy density in the fireball.
\section*{Acknowledgments}
 The authors thank Peter Braun-Munzinger and Dariusz Miskoviec
 for discussions and careful reading of the manuscript.
 This work was supported by  the Bilateral Award DLR (Germany)
- MESU (Ukraine) for the UKR 06/008 Project. The research is carried
out within the scope of the EUREA: European Ultra Relativistic
Energies Agreement (European Research Group: Heavy ions at
ultrarelativistic energies) and supported in part by the Fundamental
Researches State Fund of Ukraine: Agreement with MESU No
F33/461-2009.

\end{document}